# Overcoming the Intrinsic Performance Limitations of MEMS IMU via Diffusion-Based Generative Learning

Jiarui Lv, Feng Zhu and Xiaohong Zhang

***Abstract*—Inertial measurement units (IMUs) are fundamental sensing components in multi-source integrated navigation systems, and their performance directly determines the accuracy and reliability of solutions. However, the precision of low-cost IMUs is inherently constrained by hardware limitations. Recently, generative artificial intelligence has demonstrated remarkable capability in modeling complex data distributions and reconstructing high-fidelity signals. Motivated by this, we propose a diffusion-based generative learning framework for synthesizing high-fidelity virtual IMU data from low-cost IMU measurements. Specifically, a conditional diffusion model based on a U-Net architecture is constructed, where high-grade IMU measurements are utilized as ground-truth priors and low-cost IMU measurements are employed as conditional inputs. The virtual IMU data generated by the model is used for subsequent navigation and localization tasks. Experimental results demonstrate that the generated virtual IMU data significantly outperform the original low-cost IMU measurements in both positioning and attitude estimation. Furthermore, we transfer the model to airborne mapping experiments, where the proposed method produces thinner and more consistent point clouds. Overall, the proposed framework breaks the performance limits of low-cost IMU and demonstrates the potential of diffusion-based generative learning for virtual high-grade IMU data.**

***Index Terms*—GNSS/SINS, Diffusion Model, Generative Learning, High-Precision.**

## I. Introduction

Low-cost Micro-Electro-Mechanical System (MEMS) inertial measurement units (IMUs) have become ubiquitous in modern navigation applications, including pedestrian tracking, autonomous vehicles, drones, and wearable devices, due to their small size, low power consumption, and affordability [1], [2]. However, MEMS sensors are afflicted by significant stochastic errors, such as bias instability, angular/velocity random walk, and quantization noise, which cause unbounded drift when integrated over time in strapdown inertial navigation systems (INS) [3], [4]. This fundamental limitation severely degrades the long-term accuracy of pure inertial positioning, especially in Global Navigation Satellite System (GNSS)-denied environments [5], [6].

Traditional approaches to mitigate INS errors rely on two main streams: physical modeling of sensor errors and external aiding. Stochastic modeling techniques, such as Allan variance (AV) analysis, are widely employed to characterize the noise components of inertial sensors and construct error models (e.g., auto-regressive processes or Gauss-Markov models) that can be incorporated into Kalman filters [7], [8], [9]. Additionally, kinematic constraints derived from the platform's motion—such as zero-velocity updates (ZUPT), non-holonomic constraints (NHC), and vehicle model constraints—have been extensively used to bound error growth during GNSS outages [10], [11], [12]. While these methods provide physical interpretability, they often rely on simplified assumptions that may not hold in highly dynamic or complex environments, and their performance is limited by the fidelity of the assumed stochastic models.

In parallel, the surge of deep learning has opened new avenues for data-driven inertial navigation. Recurrent neural networks (RNNs), convolutional neural networks (CNNs), and Transformers have been applied to tasks such as IMU denoising, calibration, and direct inertial odometry [13], [14], [15], [16]. These methods can learn complex nonlinear error patterns from large datasets and often outperform traditional model-based approaches in specific conditions [1], [17]. However, purely data-driven solutions lack explicit physical consistency, which can lead to poor generalization to unseen scenarios and violations of fundamental kinematic laws [16], [18].

Recently, denoising diffusion probabilistic models (DDPMs) have emerged as a powerful class of generative models, achieving state-of-the-art results in image generation and gradually finding applications in time-series data [19]. In the context of inertial navigation, preliminary works have explored diffusion models for IMU data imputation [20] and end-to-end inertial navigation DiffusionIMU [21]. It employs an iterative denoising process to refine motion state estimates, demonstrating improved robustness to sensor noise. However, these diffusion-based approaches still operate in a purely data-driven manner and do not incorporate the rich physical knowledge available about IMU error characteristics (e.g., Allan variance parameters) or the fundamental relationships that govern inertial navigation (e.g., integration of acceleration to velocity and position).

This work was supported in part by National Science Fund for Distinguished Young Scholars of China under Grant 42425003, in part by the National Key R&D Program of China (Young Scientist Program) under Grant 2024YFB3909200, in part by the National Natural Science Foundation of China under Grant 42374031, in part by the National Natural Science Foundation of China under Grant 42388102.

Jiarui Lv is with the School of Geodesy and Geomatics, Wuhan University, Wuhan, Hubei 430079, China(e-mail: jiaruilv@whu.edu.cn).

Feng Zhu is with the School of Geodesy and Geomatics, and Hubei Luojia Laboratory, Wuhan University, Wuhan, Hubei 430079, China (e-mail: fzhu@whu.edu.cn).

Xiaohong Zhang is with the School of Geodesy and Geomatics, Hubei Luojia Laboratory, and Chinese Antarctic Center of Surveying and Mapping, Wuhan University, Wuhan, Hubei 430079, China (e-mail: xhzhang@sgg.whu.edu.cn).

To bridge this gap, we propose a diffusion-based generative learning framework that synthesizes virtual high-grade imu data from low-cost IMU measurements. The key innovations are two-fold. First, we design an Allan variance-driven, axis-dependent noise schedule that embeds sensor-specific stochastic characteristics (white noise, random walk, bias instability) into the forward diffusion process. This ensures that the reverse generative process learns to produce virtual data with realistic error dynamics. Second, we introduce two physically motivated loss functions—bias smoothness and integral consistency—that enforce the generated virtual signals to adhere to fundamental sensor physics, thereby improving both signal fidelity and navigation accuracy.

Unlike existing diffusion-based methods that focus on denoising or state estimation, our framework directly generates high-precision virtual IMU data that can be used as a drop-in replacement for raw low-cost measurements in any downstream navigation filter. This approach not only reduces positioning drift but also preserves interpretability and compatibility with existing INS/GNSS integration architectures.

The main contributions of this paper are summarized as follows:

- We propose a novel diffusion based generative learning framework that synthesizes virtual high-grade imu data from low-cost IMU measurements.
- We incorporate Allan variance-based noise modeling into the diffusion process, creating a physics-guided noise schedule that respects the distinct stochastic characteristics of gyroscopes and accelerometers.
- We introduce physical consistency constraints (bias smoothness, integral consistency) to ensure the generated virtual data adhere to kinematic principles and sensor error characteristics.

The remainder of this paper is organized as follows. Section II reviews related work on inertial navigation error modeling and data-driven methods. Section III details the proposed generative learning framework. Section IV presents experimental results and analysis. Finally, Section V concludes the paper and discusses future directions.

## II. Related Works

In this section, we review existing literature on inertial navigation error mitigation from two perspectives: traditional model-based methods and emerging data-driven approaches. We highlight their strengths and limitations to contextualize our contribution.

### *A. Traditional Error Modeling and Constraint-Based Methods*

The performance of the INS is inherently tied to the quality of its sensors. Stochastic error modeling of inertial sensors has been extensively studied to characterize and compensate for noise. Among various techniques, Allan variance (AV) has become a de facto standard for identifying noise terms such as angle/velocity random walk, bias instability, rate random walk, and quantization noise [7], [22]. The extracted parameters can be used to construct stochastic models (e.g., first-order Gauss-Markov processes) within a Kalman filter framework [8], [9]. Recent advances have extended AV to more sophisticated modeling, including generalized method of wavelet moments (GMWM) for better accuracy [9].

In parallel, kinematic constraints derived from the platform have proven effective in bounding error growth during GNSS outages. For land vehicles, non-holonomic constraints (NHC) assume zero lateral and vertical velocity, while zero-velocity updates (ZUPT) exploit stationary intervals [10], [11]. These constraints are typically fused via extended Kalman filters (EKF) or invariant extended Kalman filters (IEKF) [23], [24]. More sophisticated approaches incorporate vehicle dynamics models, such as the bicycle model or multi-layer kinematic models, to provide soft or hard constraints [11], [12]. For pedestrian navigation, biomechanical constraints of lower limbs have been exploited using multiple IMUs [25]. While these methods offer physical plausibility and real-time capability, their effectiveness relies on accurate detection of motion states and the validity of constraint assumptions, which may be violated in agile maneuvers or uneven terrains [6].

### *B. Data-driven Methods for Inertial Navigation*

The advent of deep learning has revolutionized inertial navigation by enabling end-to-end learning from raw IMU data. Early works focused on supervised learning for displacement estimation, using RNNs or CNNs to map inertial measurements to position or velocity [13], [26]. The Oxford Inertial Odometry Dataset (OxIOD) [26] provided a large-scale benchmark for training such models. Subsequently, hybrid approaches that combine model-based filters with learned components have emerged. For instance, Guo et al. [2024] integrated an IEKF with a deep network to learn noise parameters, achieving accurate dead reckoning for wheeled robots. Similarly, Or and Klein [18] proposed a hybrid filter where a neural network tunes the process noise covariance online. Other works have applied deep learning to specific tasks like IMU calibration [14], error prediction [27], and sensor fusion [15], [17].

Denoising diffusion probabilistic models (DDPMs) have recently gained traction in time-series modeling due to their ability to generate high-fidelity samples and handle complex distributions [19]. In the inertial navigation domain, Jing et al. [20] applied diffusion models to impute missing IMU data, exploiting spatiotemporal consistency constraints. Teng et al. [21] proposed DiffusionIMU, a diffusion-based framework that iteratively refines motion state estimates, achieving improved accuracy and robustness to sensor noise. However, these methods operate purely on data and do not incorporate explicit physical knowledge about sensor error characteristics or kinematic constraints. Consequently, the generated or enhanced data may not fully respect the underlying physics, limiting their utility in applications.

In summary, conventional model-based methods provide strong physical interpretability, yet their performance is constrained by model fidelity and simplifying assumptions. In contrast, data-driven approaches, particularly diffusion models, exhibit powerful representation and generative capabilities but often lack interpretability and physical consistency. To bridge this gap, this study integrates Allan variance-based noise modeling and kinematic constraints into a diffusion-based generative framework. The proposed approach enables the synthesis of virtual high-grade imu data from low-cost IMU measurements, achieving both high signal fidelity and strong physical plausibility.

## III. Diffusion-Based Generative Learning Framework for IMU Data

This section presents the proposed diffusion-based generative learning framework for synthesizing virtual high-grade imu data from low-cost IMU measurements. We first introduce the Allan variance-based stochastic error model that characterizes the physical noise properties of the target sensor (Section III-A). Next, we briefly review the fundamentals of denoising diffusion probabilistic models (Section III-B). Finally, we detail the proposed framework, including the physics-guided noise schedule, the network architecture, the physical consistency constraints, and the training/inference procedures (Section III-C).

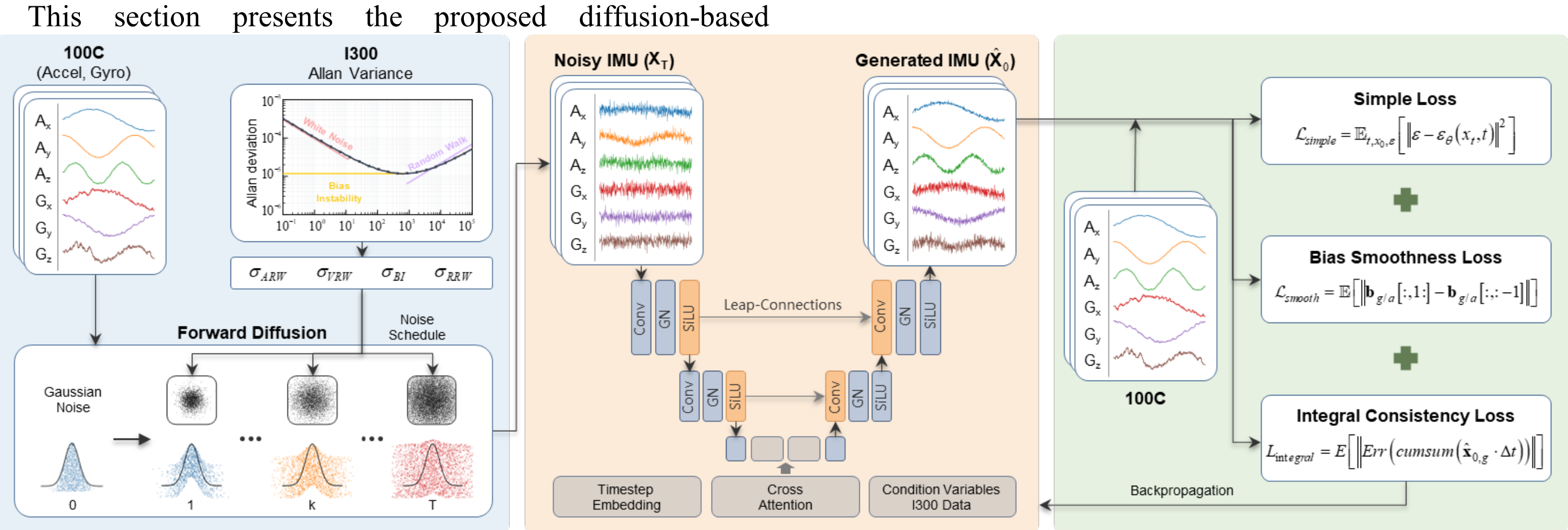

Fig. 1 Diffusion-Based Generative Learning Framework for IMU Data

### A. IMU Error Modeling via Allan Variance

Low-cost MEMS IMUs are subject to various deterministic and stochastic errors. Deterministic errors (e.g. misalignments) can be compensated through calibration, while stochastic errors are typically modeled as random processes. The Allan variance (AV) [7] is a widely adopted technique for identifying the underlying random noise components in inertial sensor data. By analyzing the sensor's output over different averaging times τ, the Allan variance plot (log-log scale) reveals distinct slopes corresponding to different noise terms. Given a static IMU recording, the Allan variance $\sigma^2(\tau)$ is computed as:

$$\sigma^2(\tau) = \frac{1}{2}\left\langle \left(\bar{\Omega}_{k+1} - \bar{\Omega}_k\right)^2 \right\rangle \tag{1}$$

where $\bar{\Omega}_k$ is the average of the sensor output over the k-th time cluster of length $\tau$. By fitting the AV curve, we extract the coefficients Quantization noise, Angle/velocity random walk, Bias instability, Rate random walk. These coefficients characterize the sensor's inherent stochastic properties and serve as priors in our diffusion model.

In our method, the raw IMU measurements are modeled as follows:

$$\begin{aligned} \hat{f}_{ib}^b &= f_{ib}^b + b_f + n_f \\ \hat{\omega}_{ib}^b &= \omega_{ib}^b + b_\omega + n_\omega \end{aligned} \tag{2}$$

where $\omega_{ib}^b$, $f_{ib}^b$ are the true angular velocity and specific force, $b_\omega$, $b_f$ are gyroscope and accelerometer bias, and $n_\omega$, $n_f$ are the gyroscope and accelerometer noises and defined as zero-mean white Gaussian noises.

$$n_\omega \in \mathcal{N}\left(0,\sigma_\omega^2\right), n_f \in \mathcal{N}\left(0,\sigma_f^2\right) \tag{3}$$

The biases are assumed to be modeled using random walk processes

$$\begin{aligned} b_\omega &= n_{\omega_b}, n_{\omega_b} \in \mathcal{N}\left(0,\sigma_{\omega_b}^2\right) \\ b_f &= n_{f_b}, n_{f_b} \in \mathcal{N}\left(0,\sigma_{f_b}^2\right) \end{aligned} \tag{4}$$

### B. Denoising Diffusion Probabilistic Models

Denoising diffusion probabilistic models (DDPMs) [19] are generative models that learn to reverse a gradual noising process. Given a clean data sample $\mathbf{x}_0 \in \mathbb{R}^d$, the forward process incrementally adds Gaussian noise over $T$ steps, producing a sequence $\mathbf{x}_1,\ldots,\mathbf{x}_T$:

$$q\left(\mathbf{x}_t \mid \mathbf{x}_{t-1}\right) = \mathcal{N}\left(\mathbf{x}_t; \sqrt{1-\beta_t}\,\mathbf{x}_{t-1}, \beta_t \mathbf{I}\right) \tag{5}$$

where $\beta_t \in (0,1)$ is a variance schedule. A convenient property is that $\mathbf{x}_t$ can be sampled directly from $\mathbf{x}_0$:

$$\mathbf{x}_t = \sqrt{\bar{\alpha}_t}\,\mathbf{x}_0 + \sqrt{1-\bar{\alpha}_t}\,\varepsilon \tag{6}$$

with $\alpha_t = 1-\beta_t$, $\bar{\alpha}_t = \prod_{s=1}^{t}\alpha_s$ and $\varepsilon \in \mathcal{N}(0,\mathbf{I})$.

The reverse process is parameterized by a neural network $\varepsilon_\theta \in (\mathbf{x}_t, t)$ that predicts the noise $\varepsilon$ added at step $t$. The training objective is a simplified denoising score matching loss:

$$\mathcal{L}_{simple} = \mathbb{E}_{t,x_0,\varepsilon}\left[\left\|\varepsilon - \varepsilon_\theta(x_t, t)\right\|^2\right] \tag{7}$$

After training, new samples are generated by starting from pure noise $\mathbf{x}_T \sim \mathcal{N}(0, \mathbf{I})$ and iteratively applying the learned reverse transitions.

### C. Diffusion-Based Generative Learning Framework for IMU Data

The proposed framework generates high-precision virtual IMU data $\hat{\mathbf{x}}_0$ from low-cost raw measurements $\mathbf{c}$ by learning the conditional distribution $p(\hat{\mathbf{x}}_0 \mid \mathbf{c})$ through a diffusion-based generative process. Unlike conventional denoising approaches that aim to remove noise, our method explicitly learns to synthesize virtual signals that emulate the behavior of a high-grade reference IMU, thereby producing data suitable for high-accuracy navigation. Fig. 1 illustrates the overall architecture. Below, we detail each component.

1) Network Architecture

The network adopts a U-Net structure with a Transformer bottleneck to capture both local and global temporal dependencies. Let $\mathbf{x}_t \in \mathbb{R}^{B\times 6\times L}$ be the noisy IMU sequence at diffusion step $t$ (batch size $B$, sequence length $L$), and let $\mathbf{c} \in \mathbb{R}^{B\times 6\times L}$ be the conditioning signal (the raw low-cost IMU measurements). The network $\varepsilon_\theta \in (\mathbf{x}_t, t, \mathbf{c})$ predicts the noise component.

a) Timestep Embedding

The diffusion step $t$ is encoded via sinusoidal positional embeddings of dimension $C$, passed through an MLP to produce a time-dependent feature vector of dimension $2C$. This vector is broadcast and added to the feature maps in the encoder.

b) Encoder

Three convolutional blocks progressively increase the channel dimension. Each block is followed by batch normalization and ReLU activation.

c) Bottleneck

A Transformer block with multi-head self-attention is applied to the $2C$ channel feature maps. It consists of multi-head attention, layer normalization, and a feed-forward network with GELU activation. This block captures long-range dependencies across the sequence.

d) Decoder

The decoder upsamples the features back to the original resolution using three convolutional blocks that gradually reduce the channel dimension from $2C$ to $C$ and finally to the original 6 channels. A skip connection concatenates the output of the first encoder block (after the initial convolution) with the bottleneck output, followed by a $1\times 1$ convolution to fuse them. This allows the decoder to access both global semantics and fine-grained local details.

2) Allan Variance Based Noise Schedule

Standard DDPMs employ a fixed scalar noise schedule $\beta_t$ (e.g., linear or cosine) that is not associated with sensor characteristics. In contrast, the proposed framework leverages the Allan variance parameters to construct an axis-dependent noise schedule that mimics the physical error accumulation of MEMS sensors.

For each axis $i$, the noise variance contributed by the random walk component grows linearly with time. We define the cumulative noise variance at an effective time $t$ as:

$$\upsilon_i(t) = \left(\sigma_{white}^i\right)^2 + \left(\sigma_{rw}^i\right)^2 \cdot t \tag{8}$$

where $t$ is normalized to the range $[0.1, 10]$ (corresponding to the diffusion steps). The axis-wise noise schedule $\beta_t^i$ is then obtained by linearly mapping the normalized variance to the interval $[\beta_{\min}, \beta_{\max}]$:

$$\beta_t^i = \beta_{\min} + (\beta_{\max} - \beta_{\min}) \cdot \frac{\upsilon_i(t) - \min_t \upsilon_i(t)}{\max_t \upsilon_i(t) - \min_t \upsilon_i(t)} \tag{9}$$

In our implementation, $\beta_{\min} = 10^{-4}$ and $\beta_{\max} = 0.02$. This schedule ensures that at early diffusion steps, white noise dominates, while at later steps, random walk accumulates, reflecting the actual error growth of the sensor.

The forward diffusion process is then performed per axis using the corresponding $\beta_t^i$. Specifically, for a clean sample $\mathbf{x}_0$, the noisy sample at step $t$ is generated as:

$$\mathbf{x}_t^i = \sqrt{\bar{\alpha}_t^i}\,\mathbf{x}_0^i + \sqrt{1 - \bar{\alpha}_t^i}\,\varepsilon^i \tag{10}$$

with $\bar{\alpha}_t^i = \prod_{s=1}^{t}\left(1 - \beta_s^i\right)$. This axis-dependent treatment respects the different noise characteristics of gyroscopes and accelerometers.

Additionally, to simulate the fact that a data window may not start from the sensor's power-on moment, we introduce a random initial bias offset $\mathbf{b}_0 \sim \mathcal{N}\left(0, diag\left(\sigma_{b_0}^2\right)\right)$ that is constant across the window. The standard deviation $\sigma_{b_o}$ is estimated from the training data by computing the standard deviation of the per-window mean differences between low-cost and reference IMU signals.

3) Physical Constraints Loss

To further enforce that the generated IMU data respects sensor physics, we incorporate three physically motivated loss terms. These losses are applied only after a warm-up period (epoch $> 10$) to avoid interfering with early training.

a) Bias smoothness loss

The gyroscope and accelerometer biases are known to vary slowly over time. We estimate the biases as the difference between the raw low-cost measurement and the generated signal in physical units:

$$\begin{cases} \mathbf{b}_{gyro} = \mathbf{x}_{gyro}^{mems} - \mathbf{x}_{0,gyro} \\ \mathbf{b}_{acc} = \mathbf{x}_{acc}^{mems} - \mathbf{x}_{0,acc} \end{cases} \tag{11}$$

The bias smoothness loss penalizes large consecutive differences:

$$\mathcal{L}_{smooth} = \mathbb{E}\left[\left\|\mathbf{b}_{g/a}[:,1:] - \mathbf{b}_{g/a}[:,:-1]\right\|\right] \tag{12}$$

b) Integral consistency loss

To suppress long-term drift, we enforce that the integral of the generated gyroscope signals matches that of the high-precision reference over the window. The integral is approximated by cumulative summation scaled by the sampling interval $\mathbf{x}_0$ :

$$\mathcal{L}_{integral} = \mathbb{E}\left[\left\| cumsum\left(\hat{\mathbf{x}}_{0,gyro}\Delta t\right) - cumsum\left(\hat{\mathbf{x}}^{gt}_{0,gyro}\Delta t\right)\right\|\right] \quad (13)$$

where $\mathbf{x}_0$ is the ground-truth gyroscope signal (from the high-grade reference sensor).

The total loss is a weighted combination:

$$\mathcal{L}_{total} = \mathcal{L}_{simple} + \lambda_1 \mathcal{L}_{smooth} + \lambda_2 \mathcal{L}_{integral} \quad (14)$$

where $\lambda_1$, $\lambda_2$ are predefined parameters.

## IV. Experiments and Analysis

This section presents the experimental setup and results of the proposed generative learning framework for high-precision virtual imu data synthesis. We first describe the dataset used, the training/test split, and the data preprocessing (Section IV-A). Then we evaluate the generation quality of the virtual data and the resulting positioning accuracy in integrated navigation (Section IV-B).

### A. Dataset Descriptions

We evaluate our method on the open-source SmartPNT-MSF dataset [28], a comprehensive multi-sensor fusion dataset designed for positioning and navigation research. The dataset was collected using two platforms (mini and mate) equipped with a variety of sensors, including multiple GNSS receivers, IMUs of different grades, optical cameras, and LiDARs. For our task of generating high-precision virtual imu data from low-cost IMU measurements, we focus on two specific inertial sensors as shown in TABLE I.

TABLE I
IMU Performance Specifications

| Sensors Type & Item | Specification |
|---|---|
| **High-precision IMU** | **Novatel SPAN-ISA-100C** |
| Gyroscope bias | $0.05°/hr$ |
| Gyroscope random walk | $0.005°/\sqrt{hr}$ |
| Accelerometer bias | $0.02mg$ |
| Measurement frequency | 200Hz |
| **MEMS-IMU** | **HGuide I300** |
| Gyroscope bias | $3°/hr$ |
| Gyroscope random walk | $0.15°/\sqrt{hr}$ |
| Accelerometer bias | $0.1mg$ |
| Measurement frequency | 200Hz |

Both sensors are rigidly mounted on the same platform, and their data are time-synchronized. The raw recordings contain timestamps and 6-axis readings (gyroscope + accelerometer) for each sensor. The dataset covers a wide range of real-world scenarios, including urban areas, campuses, tunnels, and suburban environments, offering sufficient diversity for training and testing. We selected 8 sequences for training and one separate sequence for testing, ensuring that the test trajectory is not seen during training. The detailed list is given in TABLE II. The test sequence was collected in a complex environment including tree-lined streets and urban canyons, making it a challenging benchmark.

TABLE II
Selected Data Sequences For Training And Testing

| Set | Item | Periods (s) | Scene |
|---|---|---|---|
| Train | Data01 | 2283 | ①②③ |
| Train | Data02 | 1499 | ①② |
| Train | Data03 | 2527 | ①②③ |
| Train | Data04 | 2647 | ①②③ |
| Train | Data05 | 2787 | ①②③ |
| Train | Data06 | 2494 | ①②③④ |
| Train | Data07 | 1705 | ①②③④ |
| Train | Data08 | 912 | ①②③④⑤ |
| Test | Data09 | 1783 | ①②③ |

①Open Sky, ② Urban Area, ③ Street Trees, ④ Elevated Road, ⑤ Tunnels

Fig. 2 illustrates the trajectories of all selected sequences overlaid on a satellite map. The raw IMU data are recorded at 200 Hz. We split the continuous streams into overlapping windows of length L=200 (1 second) with a stride of S=50 to augment the training data. Each window is normalized to zero mean and unit variance using statistics computed from the entire training set. The normalization parameters are saved and later applied to the test data

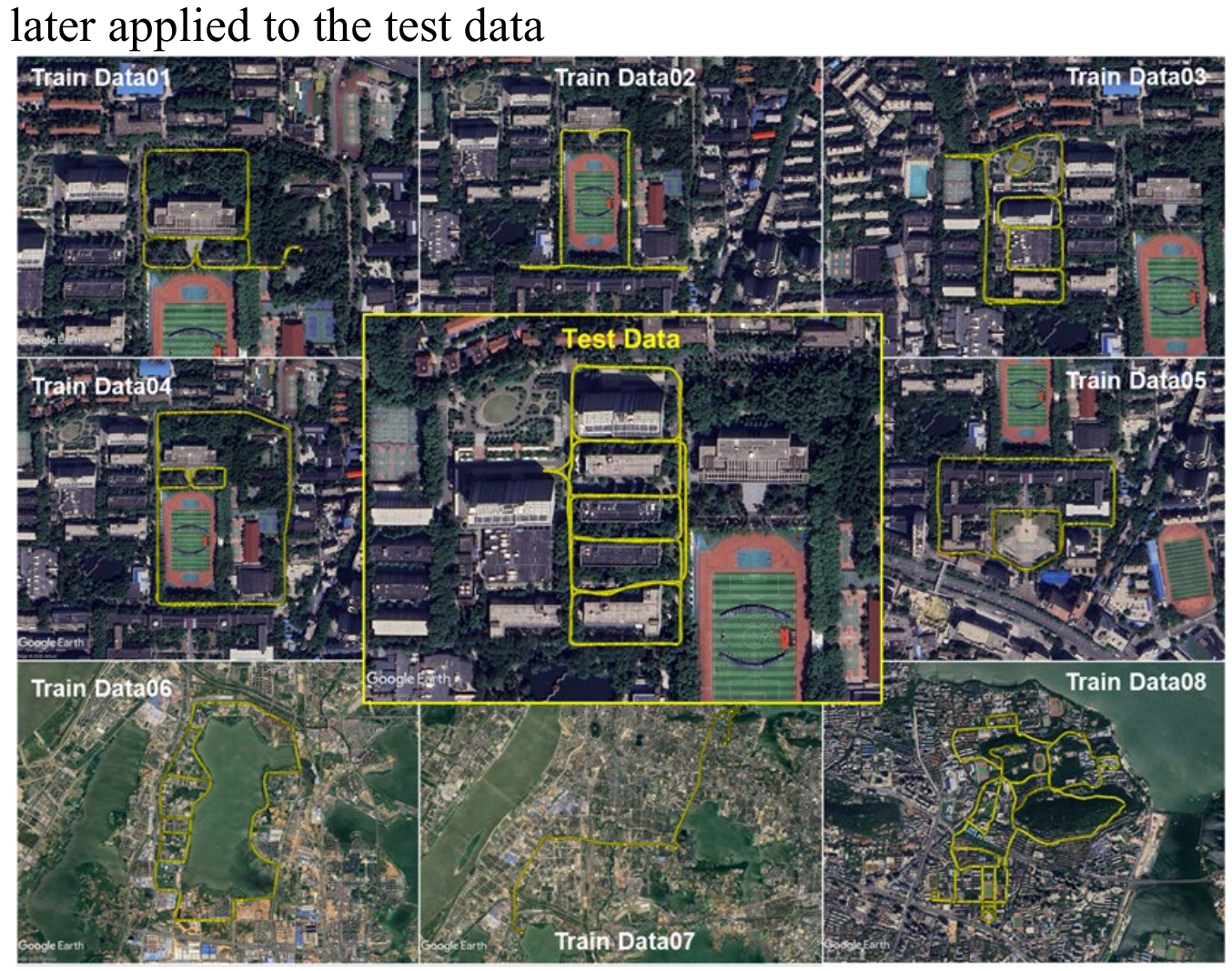

Fig. 2 Trajectories of the selected training and test sequences

### B. Generation Quality and Positioning Accuracy

To evaluate the quality of the virtual IMU data generated by the proposed method, the 100C measurements were used as the ground truth, and the I300 measurements and the generated virtual IMU data were compared. As an example, we randomly select a 0.2s IMU reading window and plot the imu data alongside the ground truth, as shown in Fig. 3. The I300 measurements (red) exhibit pronounced high-frequency noise and significant fluctuations. In contrast, the virtual IMU data generated by the proposed method (yellow) more closely follows the 100C measurements (green). This indicates that the generated virtual IMU data can more accurately characterize the true motion dynamics of the platform.

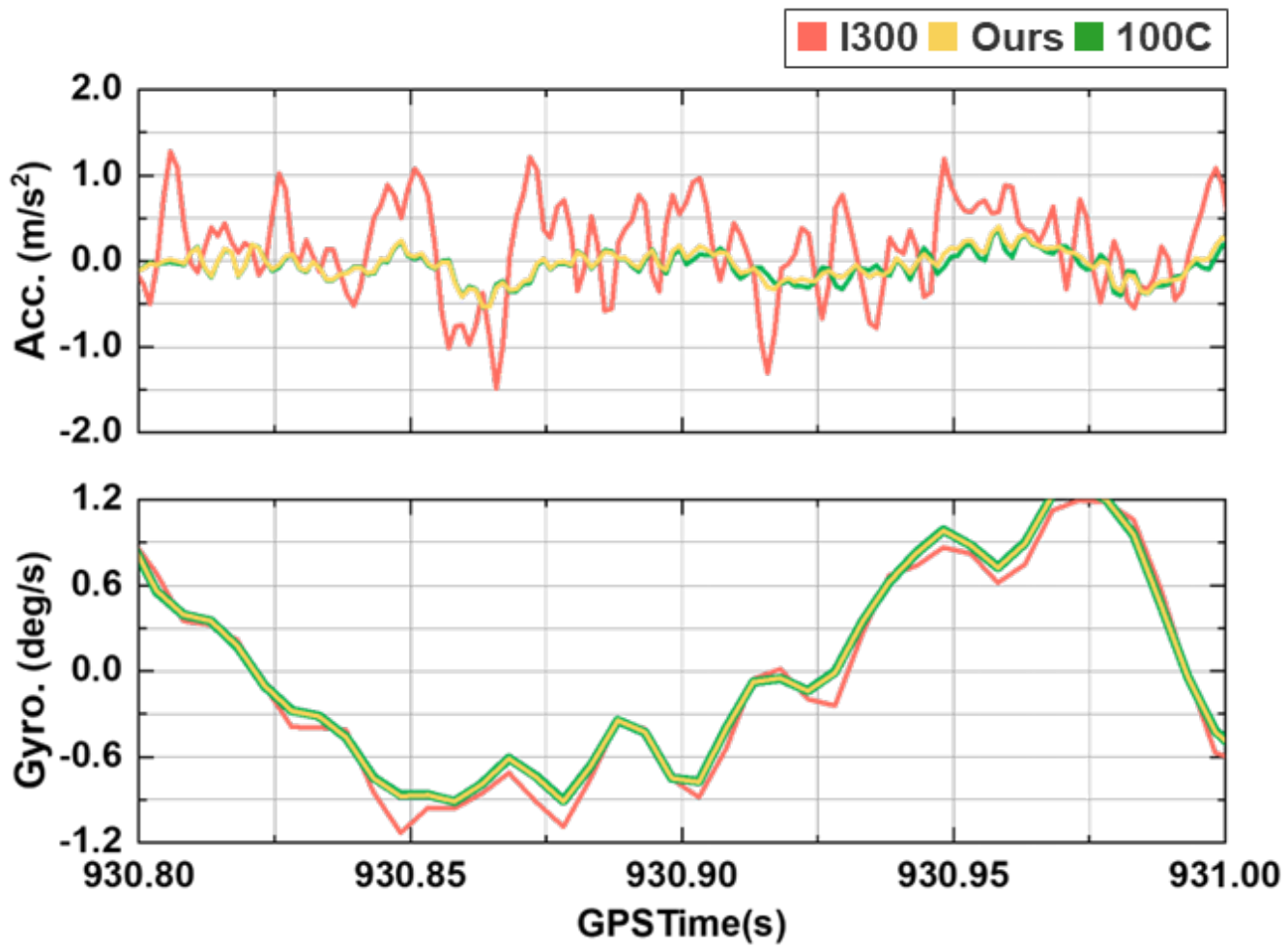

Fig. 3 Comparison of IMU measurements: I300, 100C, and generated virtual IMU data.

We further computed and compared the root-mean-square error (RMSE) of the virtual IMU data and the I300 measurements, as summarized in TABLE III. The RMSE of the proposed method was reduced by approximately 85% across six axes, further demonstrating the high accuracy and effectiveness of the generated virtual IMU data.

TABLE III
GENERATION QUALITY EVALUATION: RMSE OF IMU DATA

| Item | I300 | Ours | Improvement |
|---|---|---|---|
| Accel-X | 0.2827 | 0.0426 | 84.9% |
| Accel-Y | 1.1918 | 0.1056 | 91.1% |
| Accel-Z | 1.2723 | 0.1098 | 91.3% |
| Gyro-X | 0.1367 | 0.0202 | 85.2% |
| Gyro-Y | 0.2375 | 0.0343 | 85.5% |
| Gyro-Z | 0.2354 | 0.0333 | 85.8% |

After evaluating the performance of the proposed model, the generated virtual IMU data were further integrated into the GNSS/SINS integrated navigation system to assess their impact on positioning performance. The ground truth was obtained using the smoothed and combined solutions of multi-GNSS RTK/IMU (ISA-100C), using commercial software NovAtel Inertial Explorer (IE 8.90).

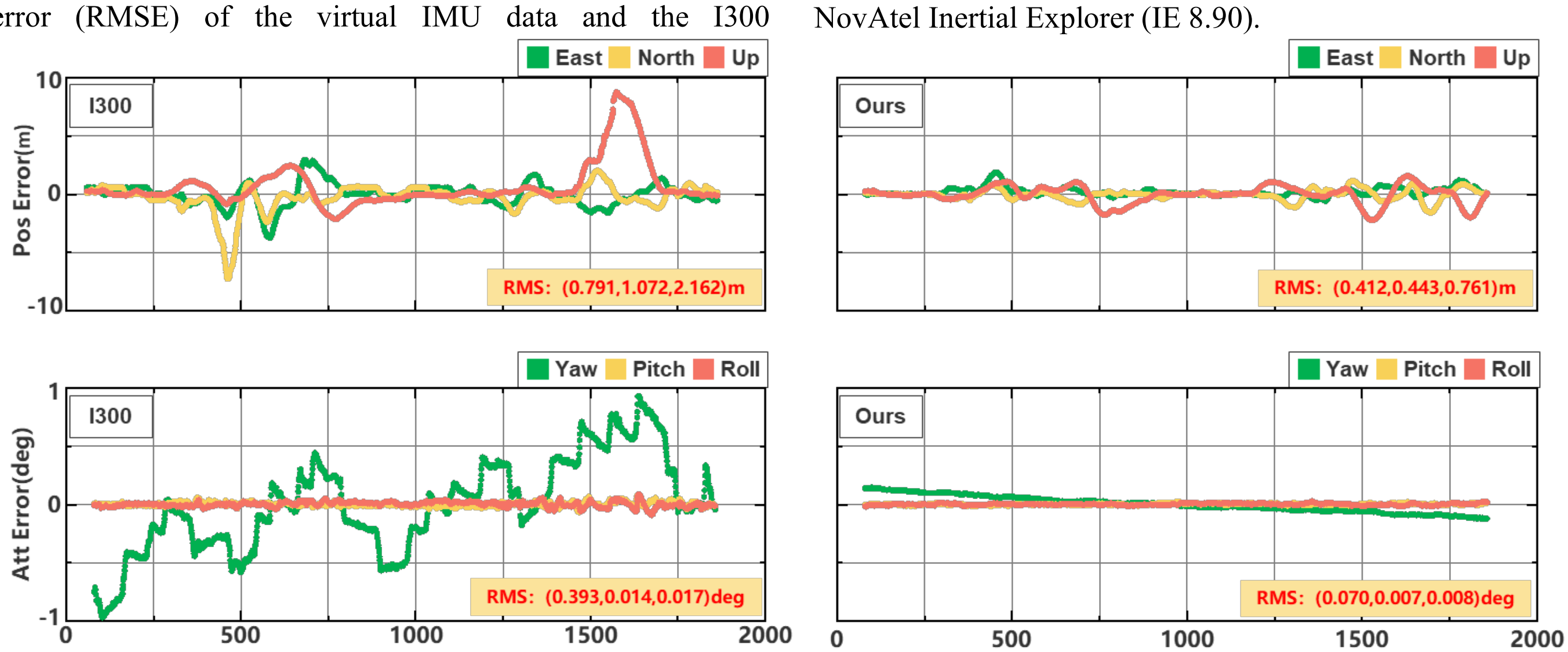

Fig. 4 Position errors (ENU) and attitude errors (yaw, pitch, roll) of the I300 measurement and generated virtual IMU data.

TABLE IV
ERROR STATISTIC: I300 VS. GENERATED VIRTUAL IMU DATA

| Statistics | Solution | E(m) | N(m) | U(m) | Position Improvement | Yaw(deg) | Pitch(deg) | Roll(deg) | Attitude Improvement |
|---|---|---|---|---|---|---|---|---|---|
| RMS | I300 | 0.791 | 1.072 | 2.162 | 61.7% | 0.393 | 0.014 | 0.017 | 82.0% |
| | Ours | 0.412 | 0.443 | 0.761 | | 0.070 | 0.007 | 0.008 | |
| MAX | I300 | 2.900 | 2.010 | 9.200 | 72.9% | 0.801 | 0.078 | 0.069 | 81.8% |
| | Ours | 1.858 | 1.118 | 1.549 | | 0.143 | 0.023 | 0.024 | |
| CEP95 | I300 | 0.963 | 0.677 | 6.157 | 75.7% | 0.575 | 0.025 | 0.024 | 78.9% |
| | Ours | 0.880 | 0.719 | 1.007 | | 0.120 | 0.012 | 0.013 | |

We comparatively analyzed the positioning errors in the east–north–up (ENU) directions, as well as the attitude errors in yaw-pitch-roll (YPR) directions, for the GNSS/SINS integrated navigation system using the I300 measurements and the generated virtual IMU data, respectively, as illustrated in Fig. 4. It can be seen that, compared with the solution using the I300 measurements, the use of the generated virtual IMU data significantly improves the performance under challenging environments. Specifically, the positioning error is reduced from 2.5 m to 0.97 m, while the attitude error decreases from 0.39° to 0.07°, demonstrating substantial improvements in both positioning and attitude estimation accuracy.

We further conducted a statistical analysis of the navigation results, as summarized in Table IV. The RMSE, maximum error, and 95% circular error probable (CEP95) of the positioning and attitude results under the two solutions were compared. Compared with the I300 measurements, the virtual IMU data generated by the proposed method improves the attitude accuracy by approximately 80% and the positioning accuracy by approximately 40%. These results demonstrate that the proposed method is capable of generating high-quality virtual IMU data, and that the improved data correspondingly enhances the accuracy and robustness of the GNSS/SINS integrated navigation system under challenging environments.

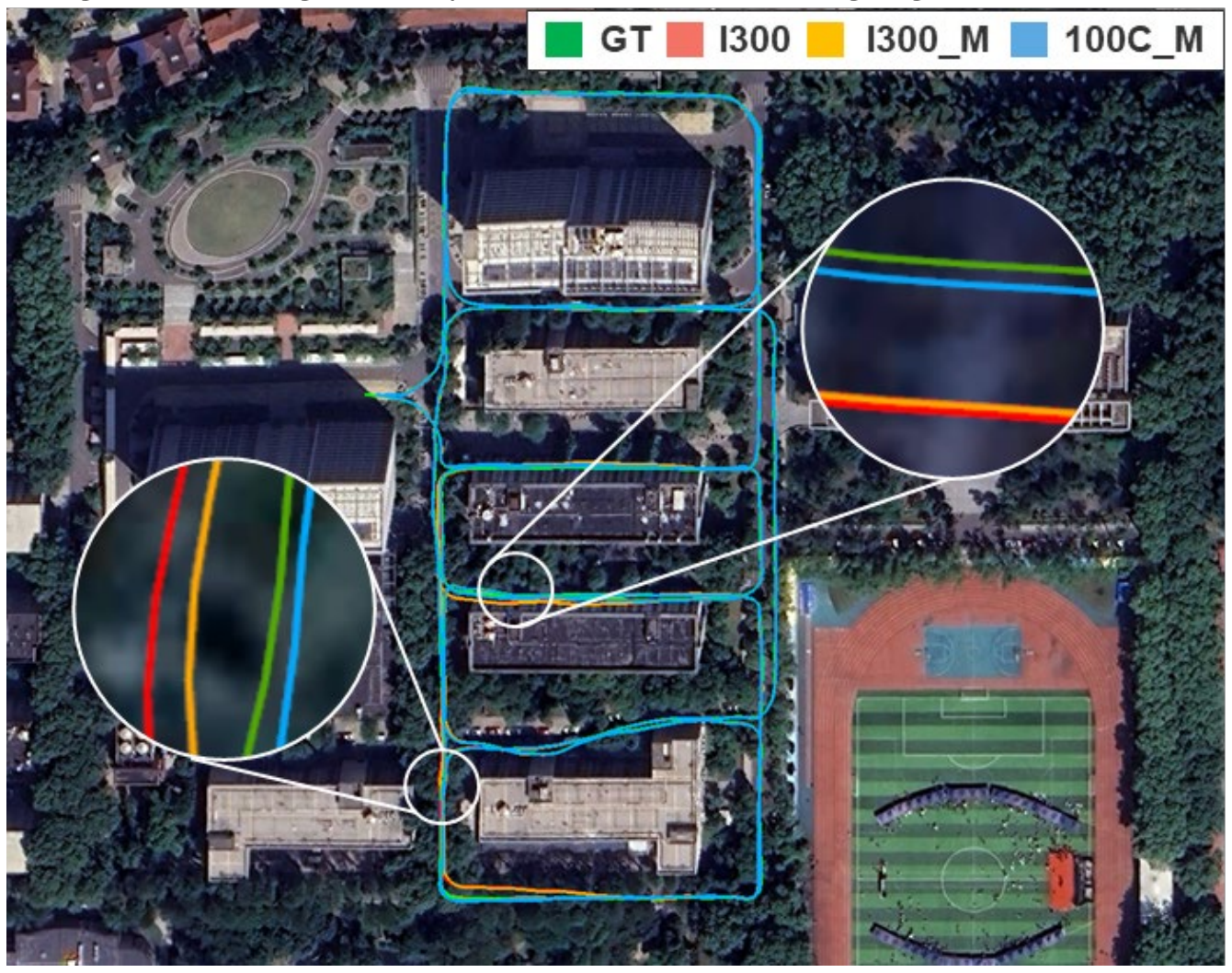

Fig. 5 Comparison of solved trajectories under different stochastic IMU error models.

During the utilization of the generated virtual IMU data, we found that, since the precision of the virtual IMU data had been enhanced, the corresponding high-precision stochastic IMU error model should also be adopted in the GNSS/SINS integrated navigation. To verify this hypothesis, comparative experiments were conducted using two different stochastic IMU error models: one employing the I300 error model and the other using the high-grade 100C error model. The corresponding navigation results were further compared with the ground truth and the original I300 measurement-based solution. The resulting trajectories are illustrated in Fig. 5.
In addition, the cumulative distribution functions (CDFs) of the positioning and attitude errors were statistically analyzed, as shown in Fig. 6. The experimental results are consistent with our expectation: the optimal performance is achieved when the generated virtual IMU data are configured with a more reasonable error model.

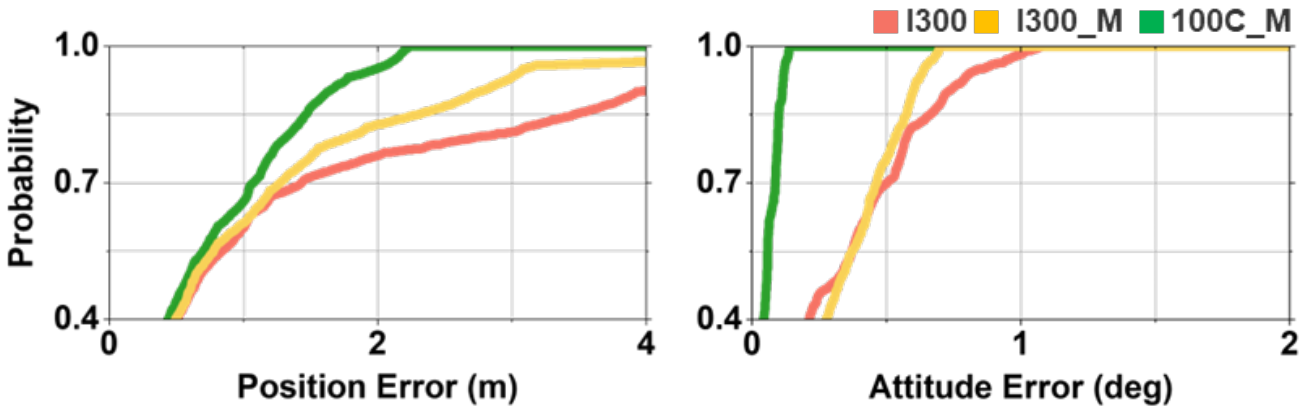

Fig. 6 Cumulative distribution of positioning and attitude errors under different stochastic IMU error models.

Finally, the proposed model was further transferred to an airborne mapping scenario to evaluate its generalization capability in practical applications. Point cloud maps constructed using the original I300 measurements and the generated virtual IMU data were comparatively analyzed, and the corresponding point cloud thicknesses were evaluated, as illustrated in Fig. 7. It can be observed that, even in relatively open environments, the point cloud map generated using the virtual IMU data is thinner, indicating improved consistency and accuracy.

These results demonstrate that the virtual IMU data generated by diffusion-based generative learning can break the performance limitations of the low-cost IMU, thereby achieving further precision enhancement in mobile mapping.

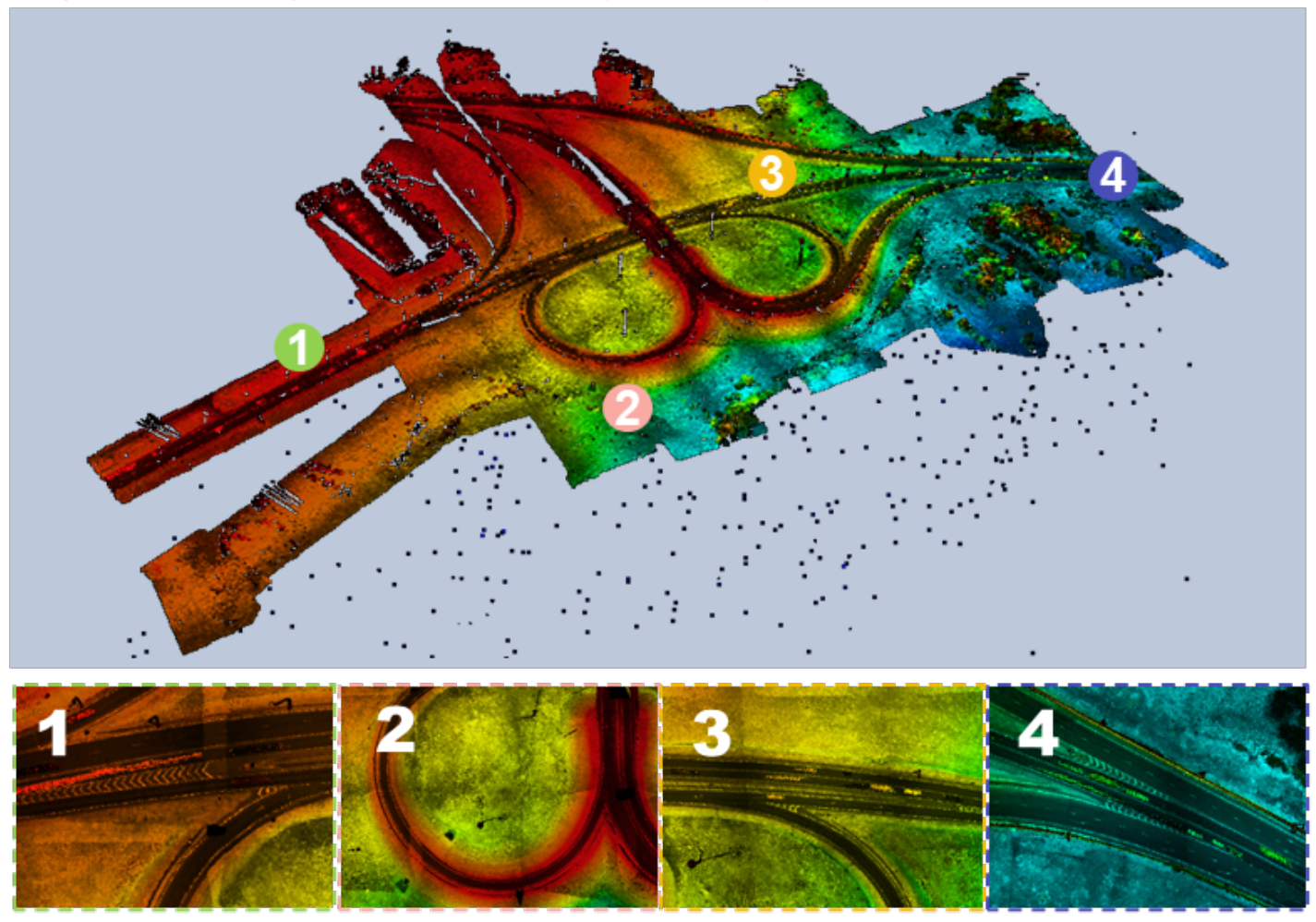

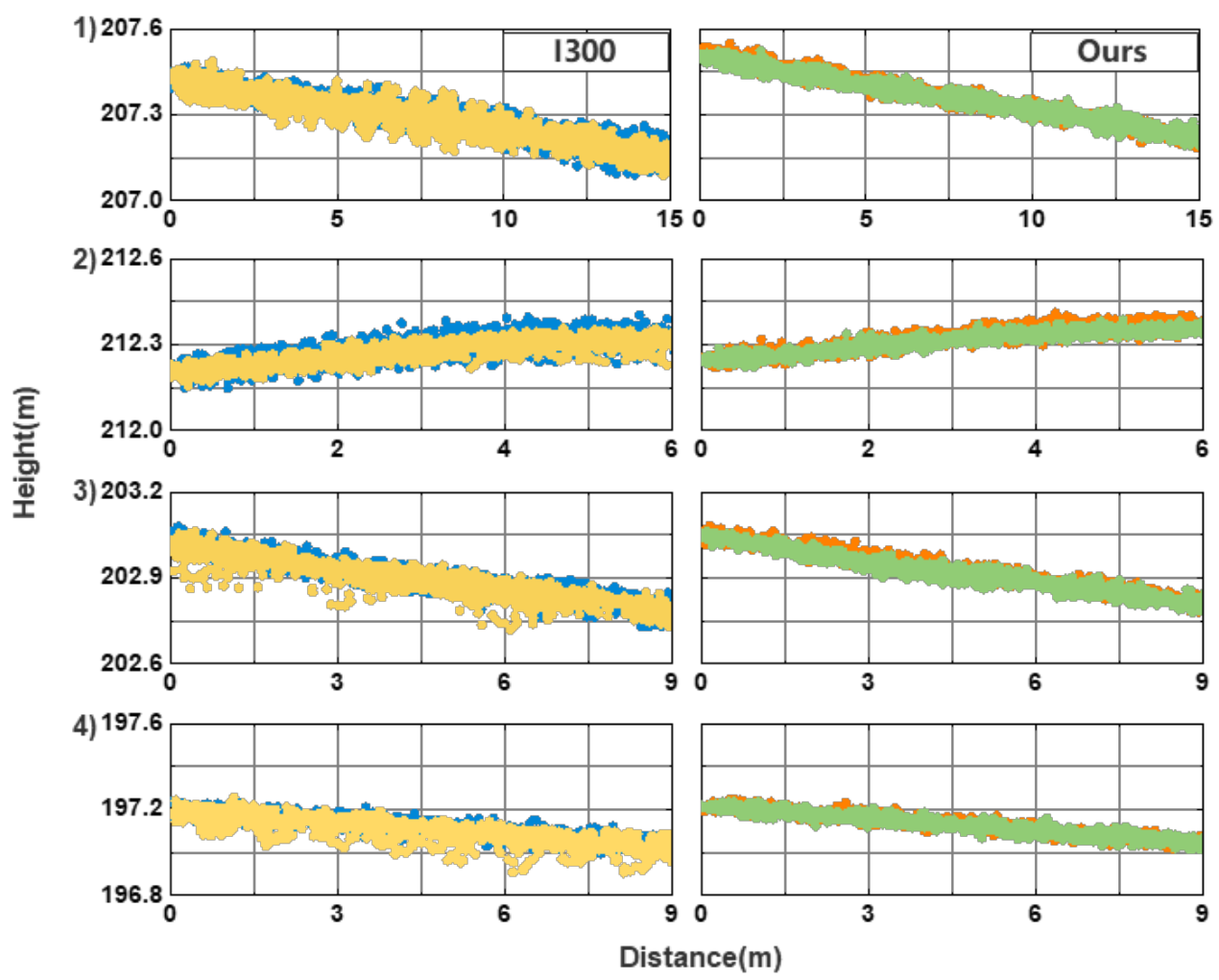

Fig. 7 Comparison of airborne point clouds generated using raw and virtual IMU data.

## V. CONCLUSION

This paper presents a diffusion-based generative learning framework for synthesizing high-fidelity virtual IMU data from low-cost IMU data. An Allan variance-guided axis-dependent diffusion strategy is developed to incorporate sensor-specific stochastic characteristics into the generative process, while physics-consistency constraints are introduced to preserve realistic inertial dynamics. In addition, a U-Net architecture with a Transformer bottleneck is adopted to capture both local and long-range temporal dependencies in inertial measurements.

Experiments on the SmartPNT-MSF dataset demonstrate that the generated virtual IMU data reduces the measurement error by approximately 85% across six axes compared with the raw I300 measurements. When integrated into the GNSS/SINS integrated navigation system, using a more reasonable stochastic IMU error model achieves even better results and higher accuracy. Furthermore, airborne mapping experiments further validate the generalization ability and effectiveness of the proposed method.

Overall, the proposed framework demonstrates the potential of diffusion-based generative learning for virtual IMU data generation, enabling precision enhancement beyond the performance limitations of low-cost IMUs. Future work will focus on multi-sensor generative fusion and real-time deployment in autonomous navigation systems.